# Giant anisotropy of spin relaxation and spin-valley mixing in a silicon quantum dot

Xin Zhang,[1,2,#] Rui-Zi Hu,[1,2,#] Hai-Ou Li,[1,2,*] Fang-Ming Jing,[1,2] Yuan Zhou,[1,2] Rong-Long Ma[1,2] Ming Ni,[1,2] Gang Luo,[1,2] Gang Cao,[1,2] Gui-Lei Wang,[3] Xuedong Hu,[4] Hong-Wen Jiang,[5] Guang-Can Guo,[1,2] and Guo-Ping Guo[1,2,6*]

[1] *CAS Key Laboratory of Quantum Information, University of Science and Technology of China, Hefei, Anhui 230026, China*
[2] *CAS Center for Excellence and Synergetic Innovation Center in Quantum Information and Quantum Physics, University of Science and Technology of China, Hefei, Anhui 230026, China*
[3] *Key Laboratory of Microelectronics Devices & Integrated Technology, Institute of Microelectronics, Chinese Academy of Sciences, Beijing 100029, China*
[4] *Department of Physics, University at Buffalo, SUNY, Buffalo, New York 14260, USA*
[5] *Department of Physics and Astronomy, University of California, Los Angeles, California 90095, USA*
[6] *Origin Quantum Computing Company Limited, Hefei, Anhui 230026, China*

[#] These authors contributed equally to this work
[*] Corresponding author. Emails: haiouli@ustc.edu.cn; gpguo@ustc.edu.cn

**Abstract**

In silicon quantum dots (QDs), at a certain magnetic field commonly referred to as the "hot spot", the electron spin relaxation rate ($T_1^{-1}$) can be drastically enhanced due to strong spin-valley mixing. Here, we experimentally find that with a valley splitting of $78.2 \pm 1.6$ μeV, this "hot spot" in spin relaxation can be suppressed by more than 2 orders of magnitude when the in-plane magnetic field is oriented at an optimal angle, about 9° from the [100] sample plane. This directional anisotropy exhibits a sinusoidal modulation with a 180° periodicity. We explain the magnitude and phase of this modulation using a model that accounts for both spin-valley mixing and intravalley spin-orbit mixing. The generality of this phenomenon is also confirmed by tuning the electric field and the valley splitting up to $268.5 \pm 0.7$ μeV.

**Main text**

Single-spin qubits in Si quantum dots (QDs) are considered one of the most promising contenders for large scale quantum computation [1-3]. In silicon, the relatively weak spin-orbit interaction (SOI) and the existence of an abundant spin-zero isotope allow the electron spin to preserve its quantum state for exceptionally long times, leading to a spin relaxation time ($T_1$) over hundreds of milliseconds [4-6] and a spin coherence time ($T_2$) over tens of microseconds [7,8]. However, adverse effects from an imperfect substrate may weaken some of these advantages [2]. In silicon QDs, the energy gap between the lowest two valley-orbit states, which are obtained by breaking six-fold degeneracy of the conduction band minima (valley), is sensitive to the interface disorder [9-12]. For spin relaxation, this energy gap, also called valley splitting, introduces a spin relaxation "hot spot" when its magnitude $E_{VS}$ matches the Zeeman energy $E_Z$ [13]. As a result, spin relaxation rate can be enhanced to $10^3$ to $10^6\,\text{s}^{-1}$ [6,14-16] depending on the environment. To mitigate such effects, it is crucial to better understand and control the interactions between the spin and valley degrees of freedom in silicon.

Over the past decade, spin relaxation in Si QDs has been investigated both experimentally [4-6,14-17] and theoretically [13,18,19]. It was found that electrical noise via SOI plays an important role in determining spin relaxation in silicon. For magnetic fields near the spin relaxation "hot spot", the relaxation process is dominated by the SOI with valley states (spin-valley mixing), while for magnetic fields away from the "hot spot", especially higher fields, $T_1$ is dominated by the intravalley SOI with higher orbital states (intravalley spin-orbit mixing). The effect of SOI on spin relaxation can be viewed as a result of an effective spin-orbit magnetic field $\boldsymbol{B_{SO}}$. A finite angle between $\boldsymbol{B_{SO}}$ and the external magnetic field $\boldsymbol{B_{ext}}$ leads to mixing of spin eigenstates [20,21], allowing electrical noises to induce spin transitions between the excited and ground states. Within this physical picture, spin mixing would vary as the angle between $\boldsymbol{B_{SO}}$ and $\boldsymbol{B_{ext}}$ is changed. Therefore, $T_1^{-1}$ should be anisotropic with respect to the external magnetic field direction.

Previous studies have revealed an anisotropic $T_1^{-1}$ in GaAs QDs [22,23] and a tunable SOI in silicon using the magnetic field direction [24,25], but so far, an anisotropic $T_1^{-1}$ in Si QDs has not been investigated. Indeed, $T_1^{-1}$ anisotropy could help improve the relaxation performance of a certain qubit by choosing an optimized magnetic field orientation. Furthermore, it is also a probe into the anisotropy of both spin-valley mixing and intravalley spin-orbit mixing.

Here, we investigate extensively the spin relaxation anisotropy near the "hot spot" in a Si metal-oxide-semiconductor (MOS) QD. We find that with $E_{VS} = 78.2 \pm 1.6$ μeV, the variation in $T_1^{-1}$ can be as large as 2 orders of magnitude at 0.8 T, but is significantly suppressed at 1.5 T. Based on a model of multiple relaxation channels and a modified picture of the effective spin-orbit magnetic field, we explain our observations by identifying the limiting mechanisms of spin-valley mixing and intravalley spin-orbit mixing. We also tune the gate voltage to examine the effect of interface electric field, and find that even if the valley splitting is increased to $268.5 \pm 0.7$ μeV by tuning the electric field, the variation range in $T_1^{-1}$ can still be up to nearly 2 orders of magnitude, with the minimal relaxation angle shifted from $8.9 \pm 0.8°$ to $1.8 \pm 2.4°$. Overall, our results should provide useful guidance for future research on spin-valley mixing and spin control experiments.

The experiment is carried out in a Si-MOS double quantum dot (DQD) device [Fig. 1(a)], though we use only one QD for $T_1$ measurements. The device is fabricated from an 8-inch natural silicon wafer grown by the float zone (FZ) method, which is near-intrinsic and has high resistivity (>10 kΩ/cm$^2$) [26]. Four layers of overlapping aluminum gates with insulating oxide in between are employed to laterally confine the QDs [26,27] (see Supplementary Material [28], Sec. 1). During the experiment, gates T, SB1 and SB2 are used to define a single electron transistor (SET) to monitor the charge state of the DQD. By differentiating the SET current $I_S$ with respect to gate voltages $V_P$ and $V_{B1}$, a charge stability diagram can be obtained [Fig. 1(b)]. Here we use (NL, NR) to refer to the number of electrons in the dot under gates P and B1, respectively, and we perform the spin relaxation measurements near the (0, 0)-(1, 0) charge transition far detuned from the interdot transition (0, 1)-(1, 0), which allows us to treat the left QD as an isolated QD. The orientation of the QD gate pattern with respect to the main crystallographic directions is also shown in Fig. 1(a) and we apply an in-plane magnetic field at an angle $\phi$ from [100] direction. For the convenience of discussion, we define [110] and [1̄10] to be the $x$ and $y$ axes, respectively.

To measure spin relaxation time $T_1$, we apply to gate P a three-step pulse sequence that was first implemented by Elzerman et al. [35], as shown by points E (empty), R (read) and W (wait) in Fig. 1 (b): first, at point W an electron is injected into the QD with a random spin state and after a time $t_{\text{wait}}$, the spin state is read out via spin-to-charge conversion by pulsing to point R, finally, the QD is emptied at point E. By measuring the spin-up probability as a function of $t_{\text{wait}}$ and fitting the data with an exponential decay, we can extract the value of $T_1$. Some examples of the exponential decays of the normalized spin-up probability $P_\uparrow$ from the experiments can be seen in Fig. 1(c), showing a striking variation in $T_1$ upon rotating the magnetic field orientation. The experimental details of the $T_1$ measurements and device parameter extraction are described in Supplementary Material [28], Sec. 2 and 3.

The measured $T_1^{-1}$ as a function of the magnetic field oriented along the direction of $\phi = 117°$ is presented in Fig. 2(a), showing a typical spin relaxation "hot spot" with $E_{\text{VS}} = 78.2 \pm 1.6$ μeV. By rotating the in-plane magnetic field orientation over the whole 360° range with a constant strength of 0.8 T and 1.5 T, we observe a sinusoidal modulation of the spin relaxation rate with a 180° periodicity. Interestingly, as shown in Fig. 2(b), while the data for the two different magnetic field strengths show a nearly common minima angle of $8.9 \pm 0.8°$ with respect to the [100] plane (see Supplementary Material, Sec. 8), the variation ranges are significantly different: for 0.8 T, $T_1^{-1}$ varies by more than 2 orders of magnitude, which is approximately 1 order of magnitude larger than that in GaAs QDs [22,23], while for 1.5 T, the variation range decreases to only six times.

To understand these distinctive behaviors of the $T_1^{-1}$ anisotropy, we first identify different origins of spin relaxation in silicon [13,15]. The expression for $T_1^{-1}$ can be written as a sum of various contributions

$$T_1^{-1} = \Gamma_{\text{J,SV}} + \Gamma_{\text{ph,SV}} + \Gamma_{\text{J,SO}} + \Gamma_{\text{ph,SO}} + \Gamma_{\text{const}}, \qquad (1)$$

where subscripts "SV" and "SO" denote spin-valley mixing and intravalley spin-orbit mixing, while subscript "J" or "ph" indicates that the type of electrical noise facilitating spin relaxation is Johnson noise or phonon noise. Different types of noise give the spin relaxation rate different power law dependences on the Zeeman energy (see Supplementary Material [28], Sec. 7) [13]. Finally, $\Gamma_{const}$ describes a relaxation channel that is independent of (or at least insensitive to) the external magnetic field [15]. By including all the major contributions to spin relaxation, we can fit the experimental data really well, and can identify the dominant relaxation channel at different field ranges, as illustrated in Fig. 2(a). In general, spin-valley mixing and intravalley spin-orbit mixing dominate spin relaxation for $\boldsymbol{B_{ext}} < 1.5$ T and $\boldsymbol{B_{ext}} > 1.5$ T, respectively, and $\Gamma_{const}$ is negligibly small for most external fields ($\boldsymbol{B_{ext}} > 0.4$ T). More specifically, for $1.5$ T $< \boldsymbol{B_{ext}} < 3$ T, $\Gamma_{J,SO}$ is much greater than $\Gamma_{ph,SO}$. Therefore, the giant $T_1^{-1}$ anisotropy at $\boldsymbol{B_{ext}} = 0.8$ T is most probably due to anisotropic spin-valley mixing, which is largely suppressed by the fast increase in $\Gamma_{J,SO}$ at $\boldsymbol{B_{ext}} = 1.5$ T. In the latter case, the anisotropy of $\Gamma_{J,SO}$ may play a role. However, since we do not observe an apparent angle shift of the anisotropy curve from 0.8 T to 1.5 T, its effect may still be negligible.

With the anisotropy of spin-valley mixing the probable cause for spin relaxation anisotropy at 0.8 T, we now examine this mechanism in more detail. It is useful to reconsider the intuitive picture of the interplay between $\boldsymbol{B_{SO}}$ and $\boldsymbol{B_{ext}}$ [20,21]. As shown in Fig. 3(a), the presence of $\boldsymbol{B_{SO}}$ causes the spin to precess around an axis different from that of $\boldsymbol{B_{ext}}$, creating a channel for the spin to relax. If $\boldsymbol{B_{SO}}$ is a real magnetic field, this spin-mixing effect would be maximum when $\boldsymbol{B_{SO}} \perp \boldsymbol{B_{ext}}$ and is zero when $\boldsymbol{B_{SO}} \parallel \boldsymbol{B_{ext}}$. As a result, the extrema position should be determined by the direction of $\boldsymbol{B_{SO}}$ and there are two opportunities in the whole rotation range for $\boldsymbol{B_{ext}}$ to be parallel or perpendicular to $\boldsymbol{B_{SO}}$, leading to a modulation cycle of 180°, which is consistent with the experimental results. However, within this simple geometric picture spin relaxation due to spin-valley mixing should be completely suppressed when the two fields are in parallel, leading to a much larger degree of anisotropy in $T_1^{-1}$, which is obviously not what we observed. To address this issue, we revisit the inter-valley spin-orbit Hamiltonian, from which $\boldsymbol{B_{SO}}$ for spin-valley mixing can be expressed as

(see also Supplementary Material [28], Sec. 6) [6,13]

$$\boldsymbol{B_{SO}} = \frac{im^*E_{VS}}{\hbar\gamma}(\alpha_m r_y^{-+}\hat{x} + \alpha_p r_x^{-+}\hat{y}). \tag{2}$$

Here, $\gamma$ is the gyromagnetic ratio, $\alpha_m = \beta - \alpha$ and $\alpha_p = \beta + \alpha$ are the SOI constants from the Dresselhaus SOI ($\beta$) and Rashba SOI ($\alpha$), and $r_y^{-+}(r_x^{-+})$ represents the intervalley dipole matrix element between the two valley eigenstates along the $y$ ($x$) axis. In general, $r_y^{-+}$ and $r_x^{-+}$ are complex numbers (see Supplementary Material [28], Sec. 6), so that the effective spin-orbit magnetic fields are also complex. To quantify the contribution of the complex terms, we introduce a complex number $R = \boldsymbol{B_{SO,x}}/\boldsymbol{B_{SO,y}} = \alpha_m r_y^{-+}/\alpha_p r_x^{-+} = Re^{i\theta}$, where $R$ is the absolute value and $\theta$ is the phase. Assuming that $\boldsymbol{B_{SO,y}}$ is fully real and $\boldsymbol{B_{SO,x}}$ is complex with a phase $\theta$, the total spin-orbit field $\boldsymbol{B_{SO}}$ can then be represented by a vector in three-dimensional space with an extra axis referring to the imaginary part of $\boldsymbol{B_{SO,x}}$ [see Fig. 3(b)], with angle $\theta$ between $\boldsymbol{B_{SO,x}}$ and the $x$ axis. A finite $\theta$ shifts $\boldsymbol{B_{SO,x}}$ away from the $x - y$ plane, so that $\boldsymbol{B_{ext}}$ in the two-dimensional plane would never be parallel to $\boldsymbol{B_{SO}}$, resulting in a residual SOI induced $T_1^{-1}$ when $\boldsymbol{B_{ext}}$ is along the minimum angle. Conversely, if the angle $\theta$ can be tuned, it would enable control of the magnitude of the spin mixing and relaxation anisotropy. Based on the parameters extracted from Fig. 2(a), a numerical calculation of $T_1^{-1}$ (see Supplementary Material [28], Sec. 7) produces a best fit with the data in Fig. 2(b) when $\theta = 3.28$ rad and $R = 1.35$. The non-zero imaginary part brought by $\theta$ leads to a reduced anisotropy of spin-valley mixing and causes a nonvanishing "hot spot" when rotating the magnetic field orientation. This can be seen by the calculated "hot spot" over the whole 360° range in the inset of Fig. 2(c). Notice other relaxation channels like $\Gamma_{J,SO}$, $\Gamma_{ph,SO}$ and $\Gamma_{const}$ cannot cause such nonvanishing "hot spot" because the "hot spot" is only determined by spin-valley mixing. In Supplementary Material [28], Sec. 7, we show that the angle of the minimal relaxation rate is also determined by the complex number $R$. It should be noted that the $8.9 \pm 0.8°$ angular deviation from [100] direction may also arise from systematic errors such as an inaccuracy in measuring the sample orientation.

However, we estimate these errors together to be no more than ±3° (see Supplementary Material [28], Sec. 2). Therefore, this deviation angle is a clear reflection of the complex nature of spin-valley mixing.

To identify the limiting mechanisms at different magnetic fields for the spin relaxation anisotropy, we numerically calculate the anisotropy magnitude $T_{1,\max}^{-1}(\phi)/T_{1,\min}^{-1}(\phi)$. As shown in Fig. 2(c), the variation range is mostly limited by $\theta$ from spin-valley mixing for $\boldsymbol{B}_{\text{ext}} < 0.85$ T, and by the residual relaxation rate $\Gamma_{J,SO}$ for $\boldsymbol{B}_{\text{ext}} > 0.85$ T. These conclusions are also illustrated in the inset of Fig. 2(b), if $\theta$ was set to $\pi$, that is, $\boldsymbol{R}$ is a real number, $T_{1,\min}^{-1}(\phi)$ would have been further reduced for $\boldsymbol{B}_{\text{ext}} = 0.8$ T, but remained nearly the same for $\boldsymbol{B}_{\text{ext}} = 1.5$ T. Notice in Fig. 2(c), the limiting mechanism of $\Gamma_{ph,SO}$ is not considered since its magnitude is much smaller than that of $\Gamma_{J,SO}$ for the range of magnetic field.

According to previous studies [6,36], the valley splitting and the valley-dependent SOI constants are dependent on the applied electric field in Si MOS QDs. Here we examine how the interface electric field affects $T_1^{-1}$ anisotropy via spin-valley mixing. As shown in Fig. 4(a), the valley splitting in our device does increase almost linearly with $V_p$ (for the measurement of the valley splitting, see Supplementary Material [28], Sec. 4). We then investigate the behavior of $T_1^{-1}$ anisotropy with $E_{VS}$ increased to $268.5 \pm 0.7$ μeV. The measured $T_1^{-1}(\boldsymbol{B}_{\text{ext}})$ along the direction of $\phi = 117°$ and $\phi = -178°$ (near the minimum $T_1^{-1}$ direction, see Supplementary Material [28], Sec 5) and the calculated "hot spot" variation by rotating magnetic field are shown in Fig. 4(b). While the "hot spot" anisotropy magnitude is similar to that in Fig. 2(c), the extrema position is shifted from $8.9 \pm 0.8°$ to $1.8 \pm 2.4°$. This can be explained by the variation of $\boldsymbol{R}$ due to the electric field change. To achieve best fit with the data, $\theta$ and $R$ in our model have to be changed to 3.36 rad and 1.1, respectively. According to previous studies, the origin of this change can be the electric field effect on the QD shape [37], SOI constants [38] or the relative position between the QD and an interfacial step [24,39]. Further insights into the electrical field effect can be obtained by independently verifying the variation of valley-dependent SOI and inter-valley transition elements. Overall, the increased electric field leads to moderate changes in both the magnitude and the orientation of $T_1^{-1}$ anisotropy, but the basic features of the giant $T_1^{-1}$ anisotropy remain even though the valley splitting is increased by over 2 times.

In the discussion above, the complex SOI field plays a significant role in determining the $T_1^{-1}$ anisotropy caused by spin-valley mixing, although the *exact* value of the SOI strength $\alpha_m/\alpha_p$ and the intervalley transition matrix elements $r_y^{-+}/r_x^{-+}$ cannot be distinguished. To extract their values, more information is needed, such as the physical mechanism of the intervalley transition elements and their dependence on the electric and magnetic fields [10,12,36,40,41]. Nevertheless, the modified picture of a complex $\boldsymbol{B_{SO}}$ mixing the spin eigenstates of $\boldsymbol{B_{ext}}$ helps us determine both the magnitude and orientation of the anisotropic spin-valley mixing, which is a clear indication that $T_1^{-1}$ anisotropy is an effective approach for characterizing spin-valley mixing in silicon. Moreover, the large anisotropy of the spin relaxation "hot spot" observed in this work also provides a method to suppress $T_1^{-1}$ in silicon QDs, which would in turn allow a larger magnetic field range for high fidelity readout and control of qubits. Such an increased workable field range may specifically inspire experiments in Si/SiGe heterostructure QDs where the valley splitting may be less controllable [15,16]. Additionally, the great modulation of spin-valley mixing may create new ways to optimize qubit performance, especially for qubits driven by spin-orbit coupling [39,42,43] (see Supplementary Material [28], Sec. 9).

In conclusion, we have studied how spin relaxation in silicon depends sensitively on the external field orientation. By rotating an in-plane magnetic field, we find that the spin relaxation rate near the spin-valley "hot spot" can be reduced by more than 2 orders of magnitude. The range of this large variation is found to be controlled both by spin-valley mixing and intravalley spin-orbit mixing. We have also shown that this great anisotropy holds in a larger electric field with slightly varied parameters of spin-valley mixing compared to the significant increase of valley splitting. For future work, the anisotropy of intravalley spin-orbit mixing at much larger magnetic fields could be investigated, which should offer a deeper understanding of the mechanism for SOI with valley and orbital states in silicon.

## Acknowledgements

We acknowledge P. Huang for helpful discussions of spin-valley relaxation. And we acknowledge D. Culcer and Z. Wang for helpful discussions of the mechanism of spin-orbit coupling. Research was supported by the National Key Research and Development Program of China (Grant No.2016YFA0301700), the National Natural Science Foundation of China (Grants No. 61674132, 11674300 and 11625419), the Strategic Priority Research Program of the CAS (Grant Nos. XDB24030601), the Anhui initiative in Quantum Information Technologies (Grants No. AHY080000), and this work was partially carried out at the USTC Center for Micro and Nanoscale Research and Fabrication. G.-L. W. acknowledges financial support by the Youth Innovation Promotion Association of the Chinese Academy of Sciences under Grant No. 2016112. H.-W. J. and X. H. acknowledge financial support by U.S. ARO through Grant No. W911NF1410346 and No. W911NF1710257, respectively.

**Figure Captions**

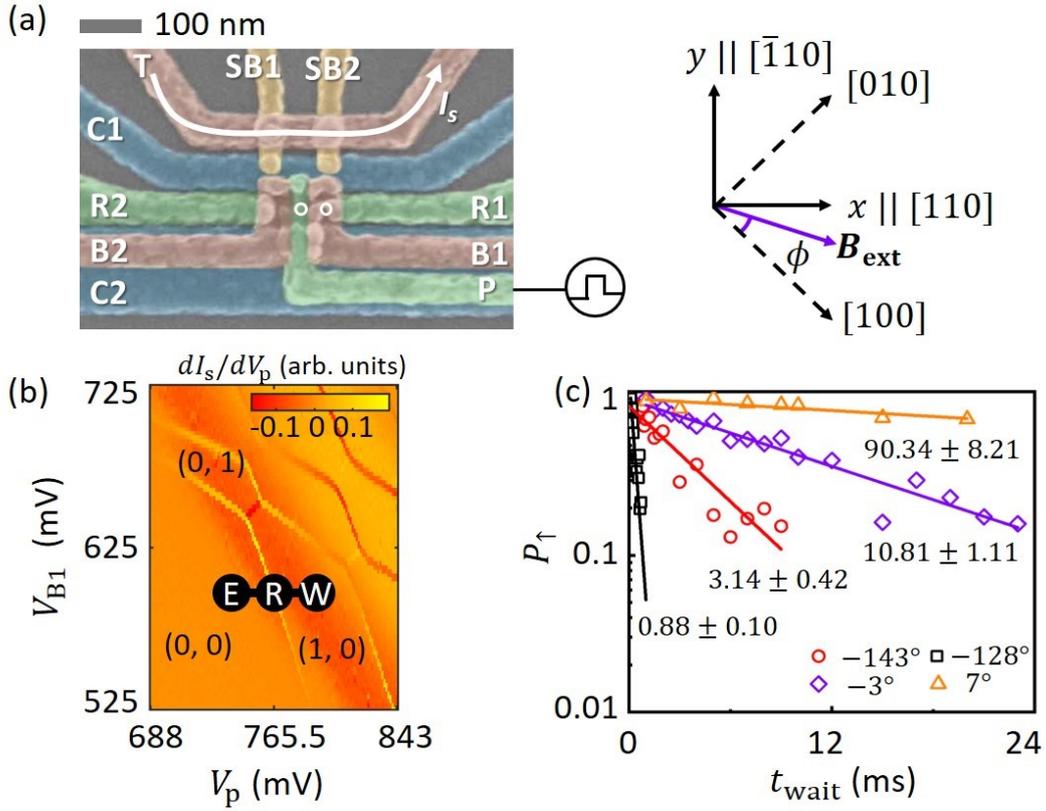

**FIG. 1.** (a) Scanning electron microscope image of a DQD device identical to the one measured. Two circles are used to proportionally denote the position and size of the dots. Inset: the crystallographic directions with respect to the sample. (b) Charge stability diagram of the DQD. The relative voltage magnitude at each step of the pulse sequence for measuring $T_1$ is overlaid on the data. (c) Normalized spin-up fraction as a function of the waiting time $t_{\text{wait}}$ for different angles $\phi$ of the 0.8 T in-plane magnetic field with $E_{\text{VS}} = 78.2 \pm 1.6$ μeV. The solid lines are exponential fits to the data with the values of $T_1$ (ms) indicated aside.

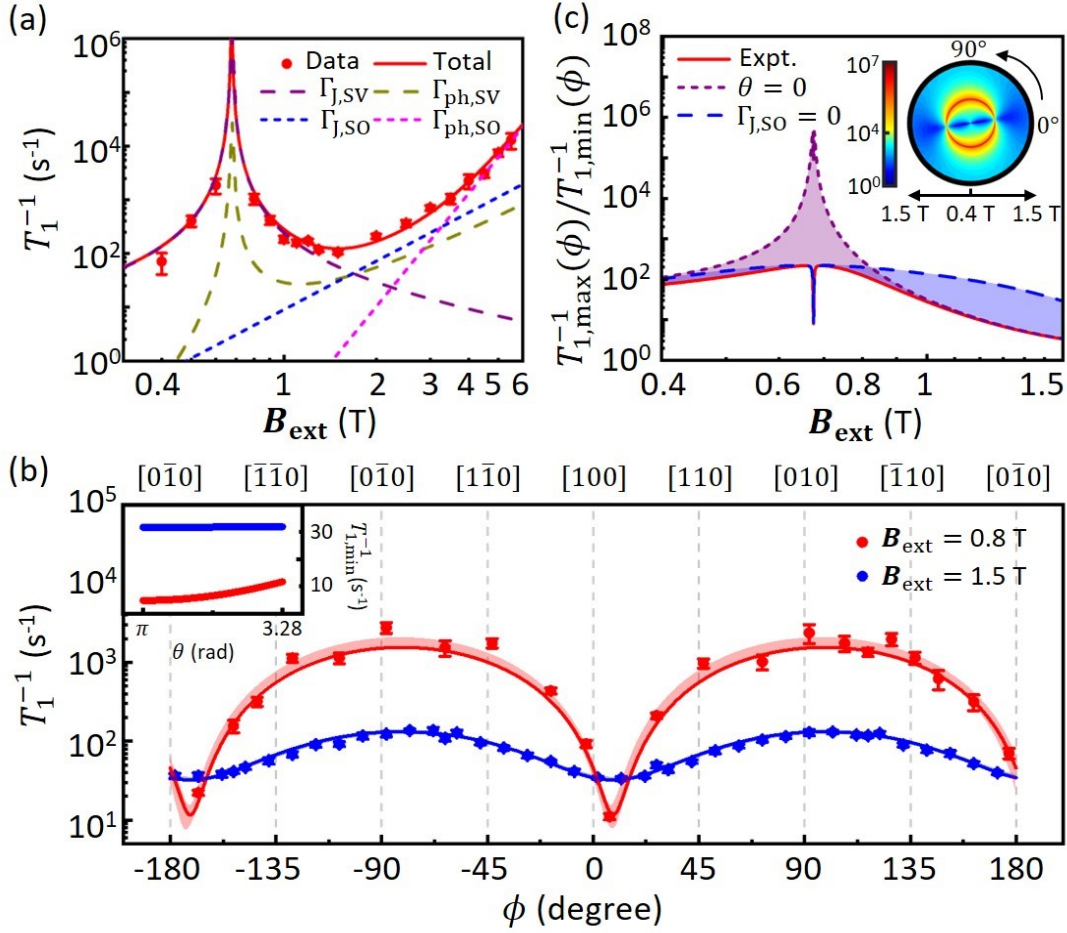

**FIG. 2.** (a) Relaxation rates as a function of the magnetic field strength with an in-plane angle of $\phi = 117°$. The fittings include contributions from different relaxation channels obtained through the model discussed in the main text. (b) Angular dependence of the relaxation rate measured with different magnetic field strengths. The red and blue solid lines are numerical results based on the spin relaxation model and the parameters from experiment, while the corresponding shaded areas indicate a 95% confidence interval with a sinusoidal fit. Inset: $T_{1,\min}^{-1}(\phi)$ as a function of the parameter $\theta$ for $B_{\text{ext}} = 0.8$ T (red) and $B_{\text{ext}} = 1.5$ T (blue). (c) Anisotropy magnitude as a function of the magnetic field strength under real experimental conditions or certain assumptions. The shaded areas indicate the amount of anisotropy suppressed by corresponding mechanism. Inset: numerical simulation of the spin relaxation "hot spot" as a function of the external magnetic field angle.

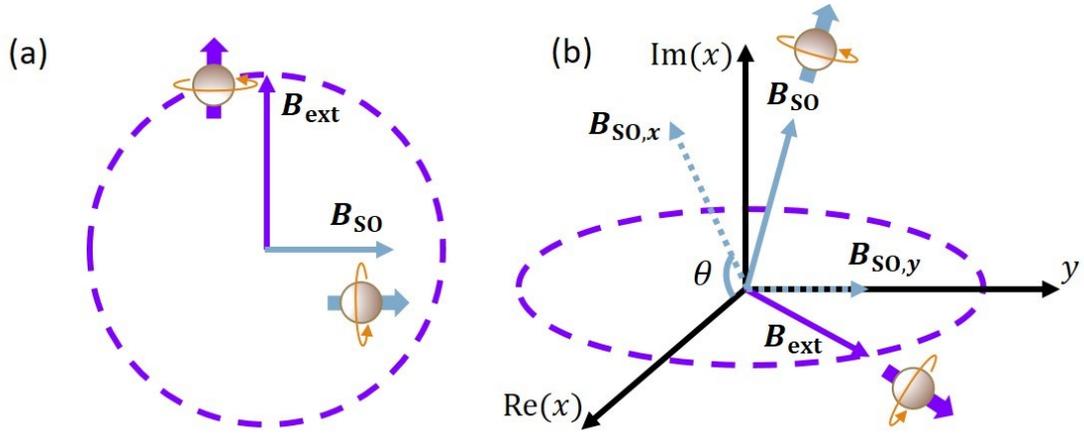

**FIG. 3.** (a) Illustration of the intuitive classical picture of the interaction between the effective spin-orbit magnetic field $B_{SO}$ and the external magnetic field $B_{ext}$. The dashed circle shows the rotation of $B_{ext}$. (b) Modified intuitive picture of the interaction between $B_{SO}$ and $B_{ext}$.

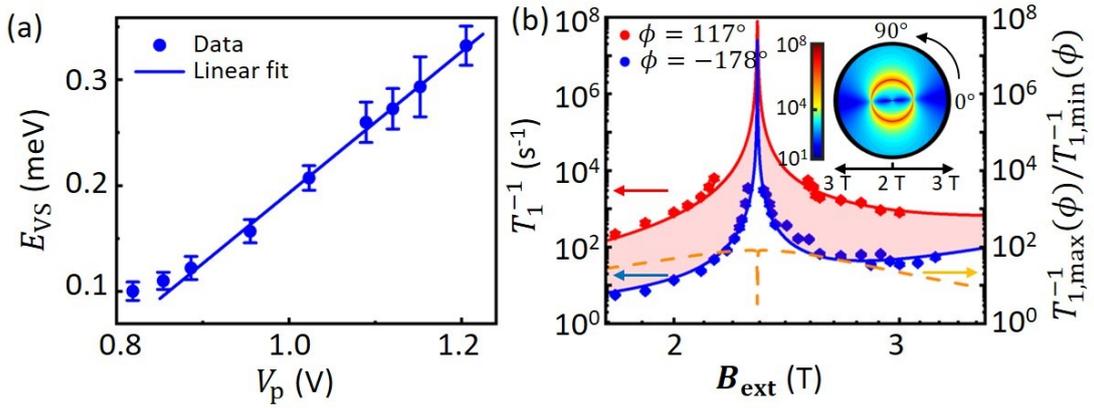

**FIG. 4.** (a) Valley splitting $E_{VS}$ as a function of the gate voltage $V_p$. A linear fit shows a tunability of $0.667 \pm 0.020$ meV V$^{-1}$. The deviation from the linear fit at small $V_p$ perhaps results from an interface localized interaction [36]. (b) Relaxation rates as a function of the external magnetic field along different directions and the calculated anisotropy magnitude with experimental parameters. Solid lines are numerical results based on our spin relaxation model and the parameters from experiment. Inset: numerical simulation of the spin relaxation "hot spot" as a function of the orientation of the external magnetic field.

# Supplementary material for "Giant anisotropy of spin relaxation and spin-valley mixing in a silicon quantum dot"

## SECTION 1: DEVICE FABRICATION

The device is fabricated on a commercially available 200 mm natural silicon wafer with a 10 nm thermally grown $SiO_2$ layer on top. First, highly doped $n^{++}$ regions are formed by P-ion implantation and Ohmic contacts are defined by 5/45 nm titanium/aurum after locally etching the silicon oxide in buffered hydrogen fluoride. To prevent leakage from gate electrodes to substrate, an additional 30 nm of alumina is then grown by atomic layer deposition on top. After that, the alumina over the active region to form quantum dots and the Ohmic contact region are locally etched. Four layers of gates (with thickness of 35/35/60/60 nm) are defined using 30 keV electron beam lithography, electron beam evaporation and liftoff of aluminum. Between each layer of the gates, a thin insulating oxide is created by oxidizing the aluminum after liftoff [1]. Finally, a forming gas anneal at 400 °C for 30 minutes is used to reduce interface traps.

The four layer of gates are shown in Fig. 1(a) in the main text: the gates C1, C2 for QD confinement and the gates SB1, SB2 for tunnel barriers of the SET constitute the first (blue) and the second (yellow) layer, respectively. Then, with an increased thickness, the gates R1, P and R2 that connect the source and drain constitute the third (green) layer. Finally, the gate T that connects the source and drain of the SET, and gates B1 and B2 that control the tunnel barriers of the QD under gate P constitute the fourth (red) layer. In principle, the second layer and the third layer could be combined. However, if we use the three-layer design, the barrier gates of the SET should be thicker, and to keep the quality of the next-layer gate (which should not be broken at the overlapped position of the SET), a much larger thickness have to be used, leading to a gate height budget of 35/60/85 nm. This gate height budget will increase the difficulty of lift-off and reduce the yield, therefore, we choose a four-layer design to obtain a relaxed height budget and a higher yield of the devices.

## SECTION 2: $T_1$ MEASUREMENT DETAILS

During the experiment, the base temperature of the dilution refrigerator (Oxford Triton) is ~ 20 mK and we estimate the electron temperature to be $180.5 \pm 8.1$ mK with a lever arm $\alpha_L \sim 0.34$ eV/V (see Section 3). For the measurements at low magnetic fields with Zeeman energy close to the electron temperature, the readout time is reduced appropriately to improve the readout fidelity [2].

Real-time detection of electron tunneling is achieved by amplifying the SET current with a room temperature low noise current amplifier (DLCPA - 200) and a JFET preamplifier (SIM910), and then low-pass filtering the amplified signal using an analog filter (SIM965) with a bandwidth of 10 kHz. Voltage pulses are applied via an Agilent 33520 signal generator. To combine the voltage pulses and d. c. voltage on the gate P, we use an analog summing amplifier (SIM 980) at room temperature. It has four input channels that can be added or subtracted with each other with a bandwidth of DC to 1 MHz and a slew rate of 40 V/μs. Its bandwidth permits both DC voltage and slow voltage pulses on the order of milliseconds and microseconds. Therefore, it is very suitable for the three-state pulse used for $T_1$ measurements in semiconductor quantum dots and free us from the capacitor problems of bias-tees.

For the inaccuracy in setting up the sample orientation, we break it into three parts: the orientation of the magnet with respect to the fridge, the orientation of the sample holder with respect to the fridge and the orientation of the sample with respect to the sample holder. For the first part, the magnet is aligned by design of the fridge company, and we confirmed the inaccuracy is well below 1° through personal communication with the Oxford Instruments Pte Ltd. For the second and the third part, they were measured optically after warming up and opening the fridge with an estimation error within ±3°. Taking into account of all the potential errors above, we estimated that the systematic error of both the in-plane and the unintentional out-of-plane magnetic field angle should be no more than ±3°. To achieve a higher accuracy for evaluating the out-of-plane angle, an experiment on Shubnikov-de Haas oscillations can be performed [2,3]. Here, since the inaccuracy is very small, we believe its effect on spin-relaxation measurements can be negligible.

For the single-shot readout of the spin state, a typical data trace is shown in Fig. S1(a), which consists of a measurement cycle E-W-R and a monitor cycle E1-W1-R1. For the measurement cycle E-W-R, it is repeated 50 – 100 times to obtain spin-up probabilities, and by changing the waiting time values, a dataset for the measurement of spin relaxation time is achieved. For the monitor cycle E1-W1-R1, it follows each measurement cycle with a fixed waiting time and is averaged after a dataset is acquired. Since the typical signal of the averaged one is already known, we can use it to judge if the measurement is implemented correctly. To ensure reliable measurements, we also calibrate the SET signal and the threshold value for single shot readout before the measurement of a new dataset. As shown in Fig. S1(b), the threshold value is obtained by analyzing an extra 1,000-shot dataset using a Gaussian mixture model. These procedures are usually repeated 20 – 100 times to get an averaged dataset for exponential fit with a fitting function. $P_\uparrow = \rho \exp\left(-\frac{t_{\text{wait}}}{T_1}\right) + \rho_0$. Fig. S1(c) shows a typical averaged dataset and an exponential fit for the measurement of a single data point in Fig. 4(b) of the main text.

For the tunneling rates of the QD to the reservoir, they are different at different stages. At stage E, we keep the tunneling-out rate of the electron ($\Gamma_{\text{out}}$) fast with respect to the empty time (5 ms) to ensure the quantum dot is emptied. A typical $\Gamma_{\text{out}}$ is about 4 kHz, which is obtained by averaging the single shot traces in Fig. S1(a) and fitting the rising edge of the empty signal with an exponential fit. At stage W, we keep the tunneling-in rate of the electron ($\Gamma_{\text{in}}$) fast with respect to the shortest waiting time (100 us) we used. A typical $\Gamma_{\text{in}}$ is much faster than 10 kHz, limited by our measurement bandwidth. At stage R, we categorize the tunnel rates into four different parts: $\Gamma_{\uparrow,\text{out}}$, $\Gamma_{\uparrow,\text{in}}$, $\Gamma_{\downarrow,\text{out}}$, $\Gamma_{\downarrow,\text{in}}$. Since the Fermi level of the reservoir is always lower than the spin-up state and higher than the spin-down state in energy, $\Gamma_{\uparrow,\text{in}}$ and $\Gamma_{\downarrow,\text{out}}$ should be very small and are negligible. Therefore, we only consider $\Gamma_{\uparrow,\text{out}}$ and $\Gamma_{\downarrow,\text{in}}$ in the following.

During the readout phase, if the electron is spin-up, it will tunnel out to the reservoir first and then an electron with spin-down tunnels back, producing a square pulse signal. As shown in the region R in Fig. S1(a), the delay time $\tau_{out}$ between the readout start and the rising edge of the pulse signal is governed by $\Gamma_{\uparrow,out}$ and the delay time $\tau_{in}$ between the falling edge and rising edge of this pulse is governed by $\Gamma_{\downarrow,in}$. By constructing a histogram of each delay time [4] from the experimental data, we could fit them into an exponential function and get a typical $\Gamma_{\uparrow,out} = 2937 \pm 424$ Hz and $\Gamma_{\downarrow,in} = 313 \pm 61$ Hz (see Fig. S2). Empirically, the tunnel rates can be affected by the readout level position, the magnetic field strength, and the magnetic field directions. In our experiment, $\Gamma_{\uparrow,out}$ varies in the range of 1 to 6 kHz, while $\Gamma_{\downarrow,in}$ varies in the range of around 30 – 300 Hz.

As the spin-relaxation rate changes as a function of the magnetic field orientation, it also affects the spin-up visibility during our experiments. In order to characterize it experimentally, we use $\rho$ in the fit function for $T_1$ to reflect the visibility of spin-up. In fact, $\rho$ is the product of spin-up readout fidelity and the probability to inject a spin-up electron. In Fig. S3, we present $\rho$ of the data in Fig. 2(b) in the main text as a function of the magnetic field orientation. It can be seen that $\rho$ varies nearly sinusoidally with an opposite trend with $T_1^{-1}$ at 0.8 T. This dependence is reasonable because during the rotation of the magnetic field, a fast spin relaxation rate should give less spin-up fraction and thus decreases the spin-up visibility with respect to the noise. This phenomenon is suppressed at 1.5 T with much smaller anisotropy of $T_1^{-1}$. Also, the higher level of $\rho$ at 1.5 T indicates higher readout fidelity due to larger Zeeman energy compared to the temperature broadening of the energy levels and longer spin relaxation time at this magnetic field.

## SECTION 3: MEASUREMENT OF LEVER ARM AND ELECTRON TEMPERATURE OF THE QUANTUM DOT

The lever arm of the measured quantum dot is extracted from the magnetospectroscopy measurements [5]. Fig. S4(a) shows the expected spin filling of the first electron into the left quantum dot when the magnetic field is applied. Assuming the g-factor to be 2, we obtain a lever arm $\alpha_L \sim 0.34$ eV/V from the slope of the charge transition. Based on the value of the lever arm, we obtain an addition energy $E_{\text{add}} = 22.48$ meV. And we estimate the dot radius $r = \frac{e^2}{8\varepsilon_r\varepsilon_0 E_{\text{add}}} = 8.6$ nm using a disk capacitor model (in the approximation of a circular quantum dot).

With a known lever arm, the electron temperature can be acquired by fitting the time-averaged quantum dot occupation as a function of gate voltage $V_P$ by a Fermi function [6]

$$N = 1/[\exp(\alpha_L(V_{P,m} - V_P)/k_B T_e) + 1], \tag{S1}$$

where $k_B$ is Boltzmann's constant, $T_e$ is the electron temperature, and $V_{P,m}$ is the gate voltage for the single-particle sate. The data and corresponding fit is shown in Fig. S4(b), with $\alpha_L = 0.34$ eV/V, and $V_{P,m}$ as well as $T_e$ being the free parameters, we obtained an electron temperature $T_e = 180.5 \pm 8.1$ mK.

## SECTION 4: MEASUREMENT OF VALLEY SPLITTING

In the main text we have shown the valley splitting $E_{VS}$ can be tuned as a function of the gate voltage, in which the exact value of $E_{VS}$ is determined from the position of the spin relaxation "hot spot". In the region where spin-relaxation hot spot appears, there should be a sudden dip of the spin-up probability, and thus the measurements can be performed quickly by observing the spin-up probability [7]. An example of the measurement is shown in Fig. S5, indicating a "hot spot" position of $2.88 \pm 0.1$ T, that is, a valley splitting energy of $333.4 \pm 11.6$ µeV.

## SECTION 5: EXPERIMENTAL RESULTS OF ROTATING THE MAGNETIC FIELD WITH A LARGE VALLEY SPLITTING

For $E_{VS} = 268.5 \pm 0.7$ µeV, we have also measured the spin relaxation rate as a function of the magnetic field orientation with a strength of 0.9 T (see Fig. S6). The reduced variation range is probably due to the smaller effect of spin-valley mixing and the residual relaxation channel $\Gamma_{\text{const}}$ that is independent of the external magnetic field. But the minimal relaxation rate angle should remain unchanged.

## SECTION 6: MODEL OF SPIN-VALLEY MIXING

In the effective mass approximation of Kohn and Luttinger and assuming relatively strong electric field at the interface, the valley wave functions can be written as [8,9]

$$|v_-\rangle = \frac{1}{\sqrt{2}}(e^{-ik_0z}u_{-z}(r) - e^{i\phi_v}e^{ik_0z}u_{+z}(r))\psi_-(r), \quad (S2)$$

$$|v_+\rangle = \frac{1}{\sqrt{2}}(e^{-ik_0z}u_{-z}(r) + e^{i\phi_v}e^{ik_0z}u_{+z}(r))\psi_+(r), \quad (S3)$$

where $k_0 = 0.85\pi/a_0$ is the position of the conduction band (also valley) minimum, $e^{\pm ik_0z}u_{\pm z}(r)$ are the Kohn-Luttinger valley functions, $\phi_v$ is the valley phase that arises from valley-orbit coupling and $\psi_\pm(r)$ are the envelop functions corresponding to the orbital ground state (s-like). For a non-ideal interface with atomic steps or roughness, $\psi_-(r)$ and $\psi_+(r)$ distinguish different envelop functions originating from valley-orbit hybridization with higher orbital states (p-like). Under this condition, the dipole matrix element $r_{-+} = \langle v_-|r|v_+\rangle$ is not zero [10] and it usually obtains a phase due to the valley phase term and the atomic scale oscillation of the Kohn-Luttinger valley functions with the sharp interface.

According to the coordinate system defined in the main text, the crystallographic direction [110] and [$\bar{1}$10] are defined as the $x$ and $y$ axes, respectively. Then the spin-orbit Hamiltonian for a 2D electron can be written as [11]

$$H_{\text{SO}} = \alpha_m P_y \sigma_x + \alpha_p P_x \sigma_y, \quad (S4)$$

where $\alpha_m = \beta - \alpha$ and $\alpha_p = \beta + \alpha$ are the constants that denote the interaction strength offered by Dresselhaus ($\beta$) and Rashba ($\alpha$) SOI contributions, $\sigma_x$ and $\sigma_y$ are the Pauli matrices along the defined axes, and $\boldsymbol{P} = \boldsymbol{p} + e\boldsymbol{A}(r)$ is the generalized momentum. Related to our experiment, the dot quantities are mainly affected by $\boldsymbol{p}$ [12] and we only consider its effect below. As mentioned in the main text, if the intervalley transition is taken into account, the new Hamiltonian should be

$$H_{\text{SV}} = \langle v_-|\alpha_m p_y \sigma_x + \alpha_p p_x \sigma_y|v_+\rangle = (im^* E_{\text{VS}}/\hbar)(\alpha_m r_y^{-+}\sigma_x + \alpha_p r_x^{-+}\sigma_y), \quad (S5)$$

in which the relationship $\langle v_-|p|v_+\rangle = (im^* E_{\text{VS}}/\hbar)\langle v_-|r|v_+\rangle$ is used.

Here, the effective spin-orbit magnetic field can thus be represented in the form of Eq. (2) in the main text, with $\boldsymbol{B}_{\text{SO},x} = (im^* E_{\text{VS}}/\hbar\gamma)\alpha_m r_y^{-+}\hat{\boldsymbol{x}}$ and $\boldsymbol{B}_{\text{SO},y} = (im^* E_{\text{VS}}/\hbar\gamma)\alpha_p r_x^{-+}\hat{\boldsymbol{y}}$. For their interaction with the external magnetic field, it can be depicted by the transition element of the spin eigenstates along the quantization axis of $\boldsymbol{B}_{\text{ext}}$, which is also called spin-valley mixing energy

$$\Delta_{SV} = 2\langle\uparrow|H_{\text{SV}}|\downarrow\rangle = 2\langle\uparrow|\boldsymbol{B}_{\text{SO}}\cdot\boldsymbol{\sigma}|\downarrow\rangle. \quad (S6)$$

Here $|\uparrow\rangle$ and $|\downarrow\rangle$ are the eigenstates of $\boldsymbol{B}_{\text{ext}}\cdot\boldsymbol{\sigma}$ and the condition for $\Delta_{SV}=0$ is equivalent to $[\boldsymbol{B}_{\text{ext}}\cdot\boldsymbol{\sigma},\boldsymbol{B}_{\text{SO}}\cdot\boldsymbol{\sigma}]=0$. Using the relationship

$$[\boldsymbol{B}_{\text{ext}}\cdot\boldsymbol{\sigma},\boldsymbol{B}_{\text{SO}}\cdot\boldsymbol{\sigma}] = 2i(\boldsymbol{B}_{\text{ext}}\times\boldsymbol{B}_{\text{SO}})\cdot\boldsymbol{\sigma}, \quad (S7)$$

this condition is further converted to $\boldsymbol{B}_{\text{ext}}\times\boldsymbol{B}_{\text{SO}}=\boldsymbol{0}$. In this context, if $\boldsymbol{B}_{\text{SO}}$ is fully real, the condition for the cross product to be zero is $\boldsymbol{B}_{\text{SO}}\parallel\boldsymbol{B}_{\text{ext}}$, while if $\boldsymbol{B}_{\text{SO}}$ is complex with a finite relative phase, the cross product would never be zero and result in a nonvanishing spin-valley mixing energy. One may notice that these conclusions are just the same as those in the modified picture of the main text.

## SECTION 7: NUMERICAL CALCULATION DETAILS

In the main text, there are five different spin relaxation channels: $\Gamma_{\text{J,SV}}$, $\Gamma_{\text{ph,SV}}$, $\Gamma_{\text{J,SO}}$, $\Gamma_{\text{ph,SO}}$ and $\Gamma_{\text{const}}$. The first two terms are about spin-valley mixing, and it can be expressed as [13]

$$T_{1,\text{SV}}^{-1} = \Gamma_{\text{J,SV}} + \Gamma_{\text{ph,SV}} = \left(c_{\text{J,SV}}\omega_Z + c_{\text{ph,SV}}\omega_Z^5\right)F_{\text{SV}}(\omega_Z), \quad (S8)$$

where $\omega_Z = E_Z/\hbar$ is the Larmor precession frequency, $c_J\omega_Z$ and $c_{\text{ph}}\omega_Z^5$ are Johnson noise term and phonon noise term, respectively, and $F_{\text{SV}}(\omega_Z) = 1 - 1/\sqrt{1+(|\Delta_{\text{SV}}|/E_{\text{VS}}-\hbar\omega_Z)^2}$ captures the extent of spin-valley mixing. The next two terms for intravalley spin-orbit mixing can be represented by two power law terms $\Gamma_{\text{J,SO}} = c_{\text{J,SO}}\omega_Z^3$ and $\Gamma_{\text{ph,SO}} = c_{\text{ph,SO}}\omega_Z^7$. The last term $\Gamma_{\text{const}}$ is a constant value and is supposed to be negligibly small in our calculation.

For numerical calculation of the $T_1^{-1}$ anisotropy, we need to know the square of the magnitude of $\Delta_{SV}$ as a function of the angle of $\boldsymbol{B}_{ext}$. Using Eq. (S6) as well as the relationship $\langle\uparrow|\sigma_x|\downarrow\rangle = i\sin\Phi$ and $\langle\uparrow|\sigma_y|\downarrow\rangle = -i\cos\Phi$, it can be expressed by [13]

$$|\Delta_{SV}|^2 = (2m^*E_{VS}/\hbar)^2[\alpha_m^2|r_y^{-+}|^2\sin^2\Phi + \alpha_p^2|r_x^{-+}|^2\cos^2\Phi -$$
$$\alpha_m\alpha_p\text{Re}(r_y^{-+}r_x^{+-})\sin 2\Phi], \tag{S9}$$

where $\Phi$ is the in-plane magnetic field angle with respect to the $x$ axis. By using the parameters from the modified picture $\alpha_m r_y^{-+}/\alpha_p r_x^{-+} = Re^{i\theta}$, Eq. (S9) could be simplified to

$$|\Delta_{SV}|^2 = c_{SV}(R^2\sin^2\Phi + \cos^2\Phi - R\cos\theta\sin 2\Phi), \tag{S10}$$

where $c_{SV}$ is a scaling factor. By substituting Eq. (S10) into Eq. (S8) and include other relaxation channels, the total equation used for numerical simulation are as follows:

$$T_1^{-1} = \left(c_{J,SV}\omega_Z + c_{ph,SV}\omega_Z^5\right)1/\sqrt{1 + (|\Delta_{SV}|/(E_{VS} - \hbar\omega_Z))^2} + c_{J,SO}\omega_Z^3 + c_{ph,SO}\omega_Z^7 +$$
$$\Gamma_{const}, \tag{S11}$$

$$|\Delta_{SV}|^2 = c_{SV}\left(\frac{1-R^2}{2}\cos(2\phi - \frac{\pi}{2}) - R\cos\theta\sin(2\phi - \frac{\pi}{2}) + \frac{1+R^2}{2}\right), \tag{S12}$$

When the magnetic field is away from the peak value of spin relaxation "hot spot", $|\Delta_{SV}|$ is much smaller than $|E_{VS} - E_Z|$, so that $F_{SV}(\omega_Z) \approx |\Delta_{SV}|^2/2(E_{VS} - E_Z)^2$, leading to $T_1^{-1} \propto |\Delta_{SV}|^2$. It is thus reasonable to regard the spin relaxation anisotropy near the "hot spot" as a direct reflection of the spin-valley mixing anisotropy, or vice versa. By calculating $|\Delta_{SV}|^2$ as a function of $R$ and $\theta$, it can help us better understand their effects on the anisotropy of both the spin-valley mixing and spin relaxation rate. Fig. S7(a) and (b) show the logarithm of the variation magnitude $\log(\Delta_{SV,max}^2/\Delta_{SV,min}^2)$ and the minimal relaxation angle $\varphi$ (with respect to the [100] axis) as a function of $R$ and $\theta$ in the complex plane of $\boldsymbol{R} = Re^{i\theta}$. When $\boldsymbol{R}$ is nearly real, the variation range is mostly dependent on $\theta$ and the minimal relaxation angle is determined by $R$, however, when $\boldsymbol{R}$ is nearly imaginary, their roles are swapped.

Notice when $R = -1$, the minimal relaxation rate angle is along [100] direction ($\varphi = 0°$), and when $R = 1$, the minimal relaxation rate angle is along [0$\bar{1}$0] direction ($\varphi = -90°$).

The parameters for numerical simulation in the main text are in the following box:

|  | Large valley splitting | Small valley splitting |
| --- | --- | --- |
| $\gamma c_{J,SV}$ | $7.2 \times 10^6$ s$^{-1}$/T | $8.5 \times 10^6$ s$^{-1}$/T |
| $\gamma^5 c_{ph,SV}$ | $1.1 \times 10^6$ s$^{-1}$/T$^5$ | $1 \times 10^6$ s$^{-1}$/T$^5$ |
| $E_{VS}$ | 269 µeV | 78.2 µeV |
| $\gamma^3 c_{J,SO}$ | 0.5 s$^{-1}$/T$^3$ | 9 s$^{-1}$/T$^3$ |
| $\gamma^7 c_{ph,SO}$ | 0.01 s$^{-1}$/T$^7$ | 0.08 s$^{-1}$/T$^7$ |
| $\Gamma_{const}$ | 0.9 s$^{-1}$ | 0.8 s$^{-1}$ |
| $c_{SV}$ | 0.02 µeV | 0.032 µeV |
| $R$ | 1.1 | 1.35 |
| $\theta$ | 3.36 rad | 3.28 rad |

Here, $\gamma$ is the gyromagnetic ratio that connects the Larmor precession frequency and the magnetic field strength. The value of $\gamma c_{J,SV}$, $\gamma^5 c_{ph,SV}$, $\gamma^3 c_{J,SO}$, $\gamma^7 c_{ph,SO}$ and $c_{SV}$ are estimated based on the results of fitting the spin relaxation "hot spot" data with Eq. S(11) without considering the angle dependence of $\Delta_{SV}$. The value of $E_{VS}$ is directly from the fitting results. Since $\Gamma_{const}$ is negligibly small, it is chosen to not affect the fitting curves. The determination of $R$ and $\theta$ is based on the numerical calculation of the variation magnitude of $\Delta_{SV}$ and minimal relaxation angle in the complex plane of $\boldsymbol{R} = Re^{i\theta}$, as shown in Fig. S7 (a) and (b), and the consideration of the fit with data.

The calculation of $|\Delta_{SV}|^2$ can also be used to analyze the effect of the out-of-plane magnetic field. Consider an external magnetic field $\boldsymbol{B}_{ext} = B_0(\sin\Theta\cos\Phi, \sin\Theta\sin\Phi, \cos\Phi)$, with the polar angle $\Theta$ and azimuthal angle $\Phi$. Using a similar derivation to that of Eq. (S10), we can obtain that

$$|\Delta_{SV}|^2 = c_{SV}\{[(R\cos\Phi + \sin\Phi)^2 + R(\cos\theta - 1)\sin 2\phi]\cos^2\Theta - 2R\sin\theta\cos\Theta + (R\sin\Phi - \cos\Phi)^2 - R(\cos\theta - 1)\sin 2\Phi\}. \text{(S13)}$$

To examine whether the out-of-plane angle $\Theta$ could help reduce $|\Delta_{SV}|^2$ to zero, we numerically calculate $|\Delta_{SV}|^2$ as a function of $\Theta$ and $\Phi$ with the experimental values of $c_{SV}$, $R$ and $\theta$. Fig. S7(c) and (d) show the minimal $|\Delta_{SV}|^2$ as a function of $\Theta$ with different parameters. As one can see, when $\theta$ is zero, both $|\Delta_{SV}|^2_{min}$ can be zero with only the in-plane magnetic field. But when $\theta$ is non-zero, an out-of-plane magnetic field is required. For Fig. S7(c), $c_{SV} = 0.032$ μeV, $R = 1.35$ and $\theta = 3.28$ rad, and a nearly zero $|\Delta_{SV}|^2_{min}$ requires an out-of-plane angle $\Theta - 90° = 3.8°$, while for S7(d), $c_{SV} = 0.02$ μeV, $R = 1.1$ and $\theta = 3.36$ rad, and a nearly zero $|\Delta_{SV}|^2_{min}$ requires an out-of-plane angle $\Theta - 90° = 6.3°$. Since these angles are out of the potential oblique magnetic field angle range ($\pm 3°$) we estimate in our device, we believe the effect of the potential oblique magnetic field is limited. It's worth mentioning that these results are from direct numerical calculation rather than analytical derivation due to the non-trivial structure of Eq. S(13). For more in-depth understanding, further theoretical research is needed.

## SECTION 8: DETAILS FOR ERROR BARS

In the main text and this supplementary material, we have mentioned many numbers with error bars, for the methods and exact value of them, we summarize as follows:

| Parameter | Value |
| --- | --- |
| Valley splitting | 78.2±1.6 μeV |
| | 268.5±0.7 μeV (average) |
| | 267.6±1.1 μeV (fit 1) |
| | 269.3±0.2 μeV (fit 2) |
| | 333.4 ± 11.6 μeV |
| Minimal relaxation angle | 8.9 ±0.8 (average) |
| | 9.8 ±0.7 (fit 1) |
| | 8.0 ±0.8 (fit 2) |
| | 1.8 ± 2.4 |
| Electron temperature | 180.5 ± 8.1 mK |
| VS tunability | $0.667 \pm 0.020$ meV V$^{-1}$ |

Note here the error bars are given by standard deviation from fitting and systematic errors are not included. The error bars of valley splitting are obtained by fitting the data of the spin relaxation hot spot with Eq. S(11) without considering the angle dependence of $\Delta_{SV}$, and the error bars of minimal relaxation angle are obtained by a simple sine-wave-curve fit of the relaxation rate as a function of the magnetic field orientation. For the data with $E_{VS} = 78.2 \pm 1.6$ μeV, there are two minimal relaxation angles that are obtained by sine-wave-curve fit of the data at two different magnetic field strength, and in the main text we represent it using the average. Similarly, for the data with $E_{VS} = 268.5 \pm 0.7$ μeV, this value is also obtained by averaging two values from fit the "hot spot" data with different magnetic field orientations. The error bars for electron temperature and VS tunability are obtained using corresponding functions.

## SECTION 9: DISCUSSION OF THE IMPLICATIONS OF THIS WORK

For the implication of anisotropic spin-valley mixing in the field of spin-orbit qubits, we take the spin-valley-orbit qubit in silicon for an example. In this system, a single spin can be driven solely by the electric field via spin-valley mixing. In the limit of $E_Z \ll E_{VS}$, the frequency of Rabi oscillation can be written as [14]:

$$\omega_R = \frac{eE_{ac}E_Z|\Delta_{SV}|r^{-+}}{\hbar E_{VS}^2}, \tag{S14}$$

where $E_{ac}$ is the applied a.c. electric field. Since $|\Delta_{SV}|$ is anisotropic, $\omega_R$ is also anisotropic. And the dephasing time ($T_2^*$) due to Stark shift can be written as [12]:

$$T_2^* = \frac{\sqrt{2}\hbar}{\Delta F_z \left|\frac{dE_z}{dF_z}\right|}, \tag{S15}$$

where $\Delta F_z$ is the standard deviation of electric field along the x axis, and $\frac{dE_z}{dF_z}$ is the derivative of Stark shift. Since $\frac{dE_z}{dF_z}$ is dependent on the external magnetic field direction, $T_2^*$ is also anisotropic. As a result, by controlling the magnetic field orientation, there should be an optimized direction to maximize the quality factor $Q = \omega_R T_2^*$ and obtain the highest control fidelity for a single qubit.

However, in a quantum dot array, the nonhomogeneity of the quantum dots could be a significant hurdle for choosing a single optimized magnetic orientation. Although the modulation phase of the spin-valley mixing anisotropy can be tuned by the electric field as demonstrated in the main text, optimizing every single qubit with the same direction is a non-trivial task. Instead of doing this, we could choose to balance the individual qubit performance and try to save the worst qubits in the array rather than improving all qubits together. In this way, we can expect an improved performance of the whole qubit array compared to the one not optimized using the anisotropy of spin-valley mixing and spin relaxation rate.

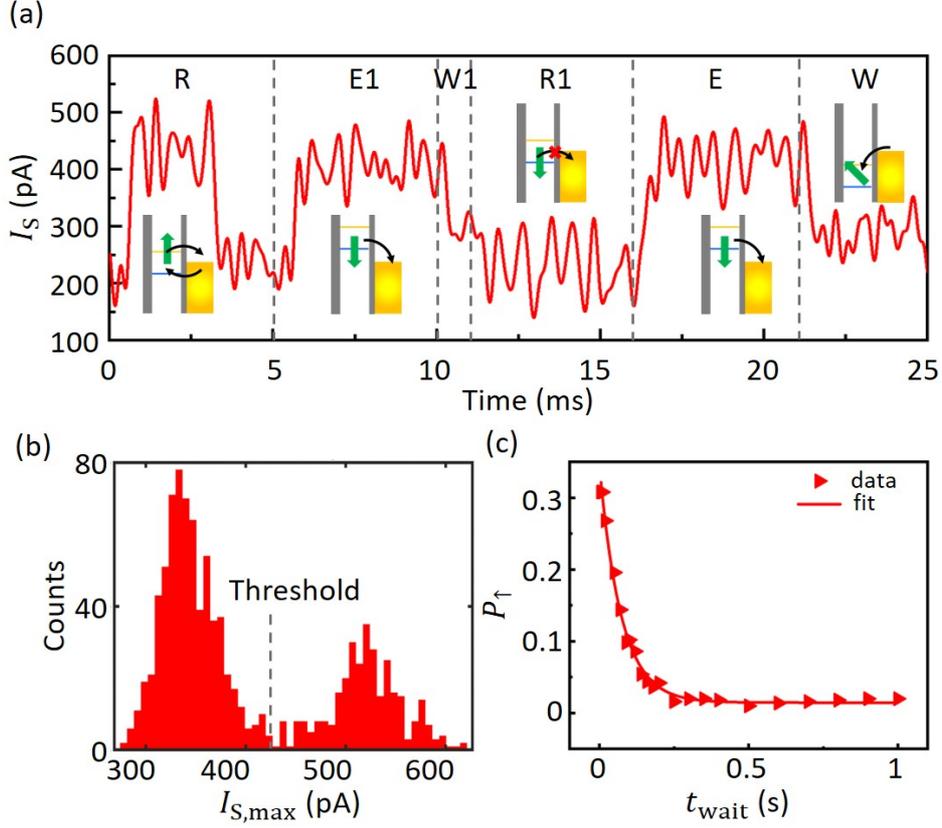

**FIG. S1.** (a) Single shot time-series of the SET current $I_S$. The corresponding measurement phase separated by gray dotted lines is indicated by R and R1 (readout), E and E1 (empty), W and W1 (wait) on the top. The energy levels in each phase are indicated by the insets near the corresponding signal except for W1, which is the same with W and is too small to put in an inset. In this signal trace, window R and R1 show the signal for the spin-up electron and spin-down electron, respectively. The delay of the signal in each window is due to the finite tunnel time of the electron. (b) Histogram of the maximum values of $I_S$ in the readout window from a 1,000-shot dataset. The threshold obtained from Gaussian mixture model is 424.41 pA. (c) An example trace of the spin-up fraction $P_\uparrow$ as a function of the waiting time $t_{wait}$ at $B_{ext} = 2$ T with an in-plane angle $\phi = -178°$. An exponential fit to it gives $T_1 = 73.3 \pm 2.9$ (ms).

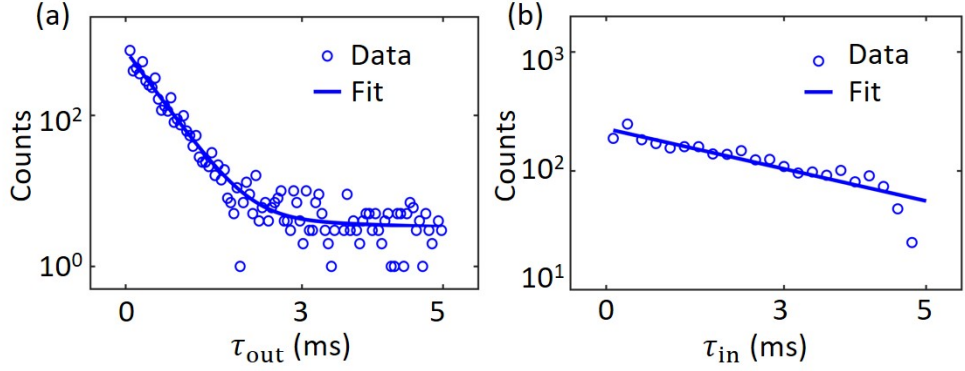

**FIG. S2.** Measurement of the tunneling rate (a) $\Gamma_{\uparrow,\text{out}}$ and (a) $\Gamma_{\downarrow,\text{in}}$ during the readout phase with exponential fits of the data.

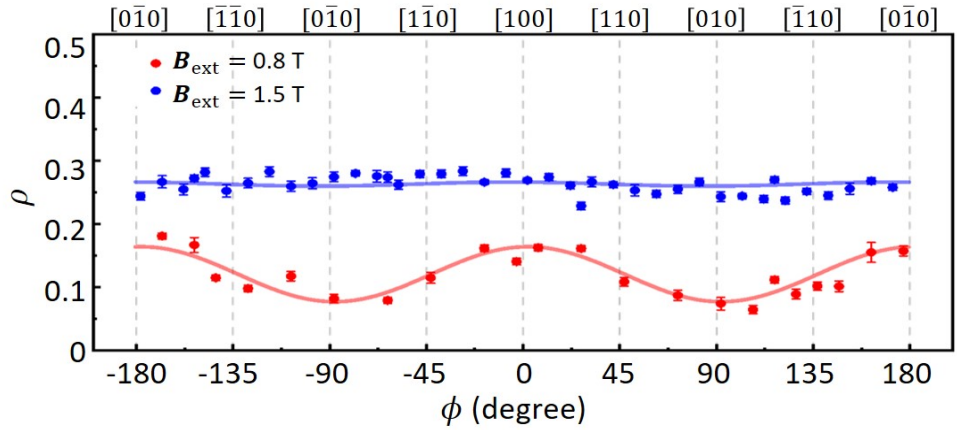

**FIG. S3.** Angle dependence of $\rho$ of the data in Fig. 2(b) in the main text. Both datasets are fitted by sine function.

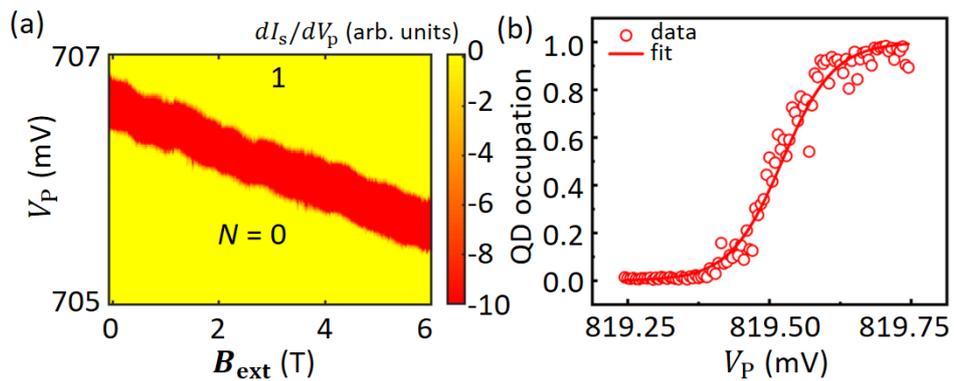

**FIG. S4.** (a) Magnetospectroscopy of the first electron filling the quantum dot. (b) Time-averaged quantum dot occupation as a function of the gate voltage $V_\text{P}$. The red solid line shows the Fermi function fit.

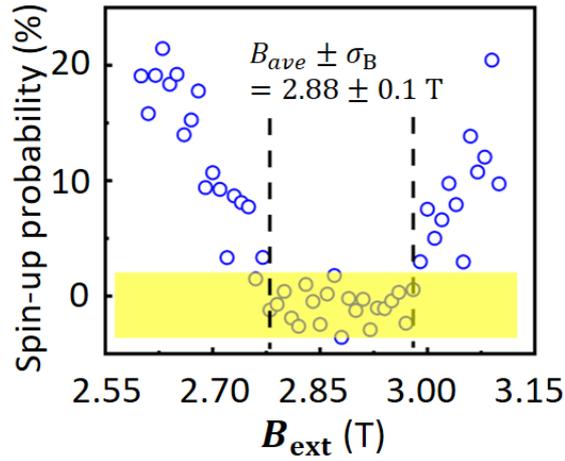

**FIG. S5.** Spin-up probability as a function of the magnetic field strength for $V_P = 1.21$ V. A sudden dip of the spin-up probability allows us to extract the "hot spot" position $B_{ave} \pm \sigma_B$, and in turn the exact value of valley splitting.

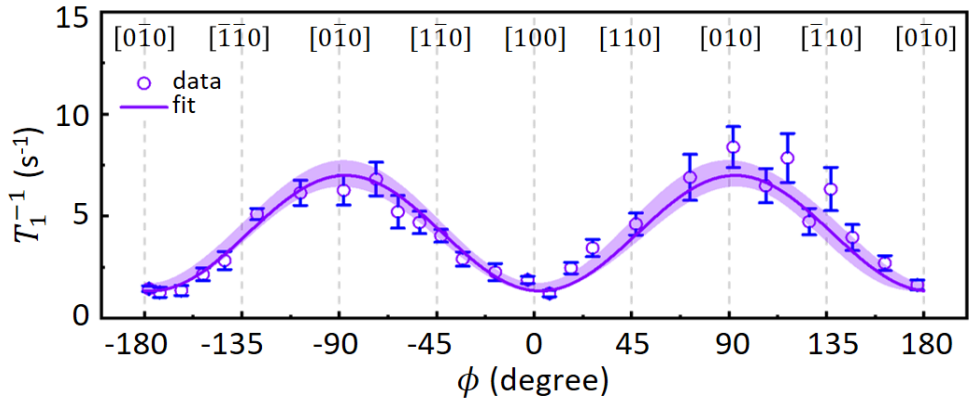

**FIG. S6.** Angle dependence of the relaxation rate measured with a 0.9 T in-plane magnetic field for $E_{VS} = 268.5 \pm 0.7$ µeV. A sine-wave-curve fit of the data reveals a minimal relaxation angle $\varphi = 1.8 \pm 2.4°$. The shaded area indicates 95% confidence interval with a sine-wave-curve fit.

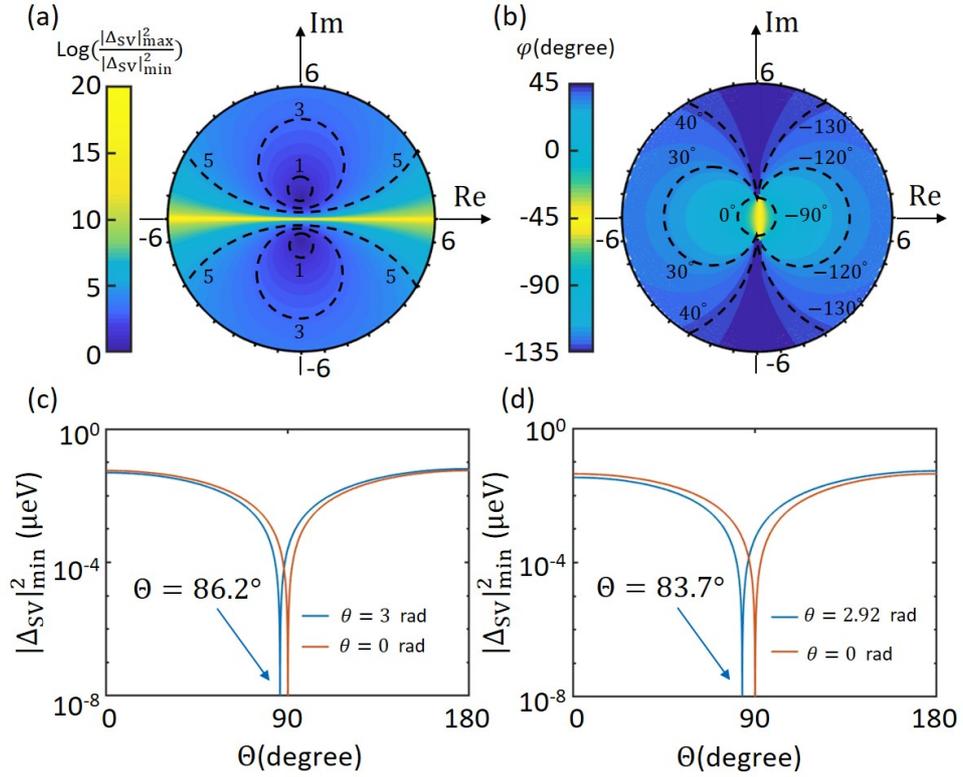

**FIG. S7.** (a) and (b) are the numerical simulations of the logarithmic variation magnitude $\log(|\Delta_{SV}|^2_{max}/|\Delta_{SV}|^2_{min})$ and the minimal relaxation angle $\varphi$ in the complex plane of $\boldsymbol{R} = Re^{i\theta}$. (c) and (d) are the numerical simulations of the $|\Delta_{SV}|^2_{min}$ as a function of the polar angle $\Theta$ with different parameters from experiments. Note $\Theta = 90°$ refers to an in-plane magnetic field.